\documentstyle[aps,prl,floats,psfig]{revtex}
\input psbox.tex	
\begin{document}

\title{Quantum critical exponents of a planar antiferromagnet
\footnote{To be published in Computer Simulations in Condensed Matter Physics
X, ed. D.P. Landau et al., (Springer Verlag, Heidelberg, 1997).}}
\author{Matthias Troyer and Masatoshi Imada}
\address{
Institute for Solid State Physics, University of Tokyo, Roppongi
7-22-1, Tokyo 106, Japan }

\maketitle

\begin{abstract}
We present high precision estimates of the exponents of a quantum
phase transition in a planar antiferromagnet. This has been made
possible by the recent development of cluster algorithms for quantum
spin systems, the loop algorithms. Our results support the conjecture
that the quantum Heisenberg antiferromagnet is in the same
universality class as the O(3) nonlinear sigma model. The Berry phase
in the Heisenbrg antiferromagnet do not seem to be relevant for the
critical behavior.

\end{abstract}
 
\section{Introduction}

Instead of classical transitions controlled by temperature $T$ a
quantum phase transition between a symmetry broken phase with
long-range N\`eel order and a quantum disordered state with
a finite spin excitation gap may be realized at $T=0$ by controlling a
parameter $g$ to increase quantum fluctuations. Criticalities around
such quantum phase transitions at $g=g_c$ may reflect inherent quantum
dynamics of the system and yield unusual universality classes with
rich physical phenomena.

The most prominent example are the high temperature
superconductors. There the quantum spin fluctuations are thought to
lead to $d$-wave superconductivity as soon as antiferromagnetism is
suppressed by hole doping. This close connection between
antiferromagnetism, quantum fluctuations and high temperature
superconductivity has triggered many theoretical investigations.

Most of these investigations are based on a mapping to an effective
field theory, the 2D O(3) quantum nonlinear sigma model. This sigma
model is in the same universality class as the 3D O(3) classical sigma
model or the 3D classical Heisenberg model.  A large number of
detailed predictions about quantum critical behavior has been made for
the sigma model \cite{CHN,CSY}. However the spin-1/2 quantum
antiferromagnet generally contains Berry phase terms \cite{haldane}
that are not present in the sigma model. The relevancy of these terms
is not clear.

In order to shed light onto this question we have simulated a two
dimensional quantum antiferromagnet (2D QAFM) that exhibits a quantum
phase transitions. We calculate the critical exponents to determine
the universality class and to check predictions made based on the
nonlinear sigma model. First results have been published in
Ref. \cite{letter}.

\subsection{Quantum critical exponents}
The critical exponents of a quantum phase transition at $T=0$ can be
defined similar to a classical finite temperature phase
transition. The quantum mechanical control parameter $g$ plays the
role of the temperature in the classical system. Approaching the
quantum critical point from the disordered side ($g>g_c$) the
correlation length diverges as
\begin{equation}
\xi\propto(g-g_c)^{-\nu}.
\end{equation}
By the Trotter-Suzuki mapping the $d$-dimensional quantum system can
be mapped onto a $d+1$-dimensional classical system. At zero
temperature the system is infinite also in the additional imaginary
time direction. The space and time dimensions are however not
necessarily equivalent, and the correlation length in the time
direction diverges in general with a different exponent
\begin{equation}
\xi_{\tau}\propto(g-g_c)^{-z\nu},
\end{equation}
where $z$ is the dynamical exponent. In a Lorentz invariant system
space and time directions are equivalent and $z=1$.
Related to the divergence of the correlation length
is a vanishing of the spin excitation gap
\begin{equation}
\Delta\propto(g-g_c)^{z\nu}.
\end{equation}
When passing through the critical point long range order is established.
The order parameter in the case of a N\'eel ordered antiferromagnet is the
staggered magnetization
\begin{equation}
m_s = \left|{1\over N}\sum_{\bf r} e^{i{\bf Qr}}S^z_{\bf i}\right|^2,
\end{equation}
where $N$ is the number of spins in the lattice, $S^z_{\bf r}$ the
$z$-component of the spin at site $\bf r$ and ${\bf Q}=(\pi,\pi)$.
Close to the critical point the staggered magnetization behaves as
\begin{equation}
m_s\propto (g_c-g)^{\beta}.
\end{equation}

At the critical point itself the real space correlation show a
power-law falloff
\begin{equation}
\label{eq:eta}
\langle S^z_{\bf 0}S^z_{\bf r}\rangle \propto 
e^{i{\bf Qr}} r^{-(d+z-2+\eta)},
\end{equation}
where $\eta$ is the correlation exponent. These three exponents are
related by the usual scaling law
\begin{equation}
\label{eq:scaling}
2\beta = (d+z-2+\eta)\nu,
\end{equation}
where the effective dimension is $d+z$ in a quantum system.

\subsection{Predictions from the nonlinear sigma model}

As mentioned above most analytic calculations of quantum critical
behavior are based on the 2D O(3) quantum nonlinear sigma model
(QNL$\sigma$M). Here we want to review the critical properties of the
sigma model relevant for the current study.
\begin{table}
\caption{Critical exponents $\beta$, $\nu$, and $\eta$.
Listed are both the estimates without making any assumption for $z$, and the
best estimate if Lorentz invariance ($z=1$) is assumed.
For comparison the exponents of the 3D classical Heisenberg (O(3)) model,
the 3D Ising model and the 2D quantum mean field exponents are listed.}
\begin{tabular}{l|cccc}
model & $\nu$ & $\beta$ & $\eta$ \\
\hline
2D QAFM & $ 0.685 \pm 0.035 $ & $ 0.345 \pm 0.025 $ & $ 0.015 \pm 0.020 $ \\
Lorentz invariant 2D QAFM & $ 0.695 \pm 0.030 $ & $  0.345 \pm 0.025 $ & $ 0.033 \pm 0.005 $ \\
\hline
3D O(3) \protect{\cite{expO3}} & $ 0.7048 \pm 0.0030 $ & $ 0.3639 \pm 0.0035 $ & $ 0.034 \pm 0.005 $ \\
3D Ising \protect{\cite{expising}} & $ 0.6294 \pm 0.0002 $ & $ 0.326 \pm 0.004 $ & $ 0.0327 \pm 0.003 $ \\
mean field & 1 & 1/2 & 0 \\
\end{tabular}
\end{table}

The critical exponents of the QNL$\sigma$M can be determined from
simple symmetry, universality and scaling arguments \cite{CHN,CSY}.
Lorentz invariance implies that $z=1$.  Furthermore the 2D QNL$\sigma$M
is equivalent to the 3D classical sigma model.  This in turn is in the
universality class of the 3D classical O(3) model, or the classical 3D
Heisenberg ferromagnet. The exponents $\beta$, $\nu$ and $\eta$ should 
thus be the same as the well known classical exponents of these models
(see Tab. I).

Chakravarty, Halperin and Nelson have discussed the phase
diagram of a planar Heisenberg antiferromagnet in the framework of the
QNL$\sigma$M.  They concentrate on the ordered phase and describe it
as a classical 2D antiferromagnet with renormalized parameters.

\begin{table}
\caption{Universal prefactor $\Omega_1(\infty)$ in the linear temperature
dependence of the uniform susceptibility at criticality. Listed are
the results for the quantum nonlinear sigma model in a $1/N$
expansion, the results by classical Monte Carlo simulation on a 3D
classical rotor model and the result of the present study.}
\begin{tabular}{ll|c}
method & Ref. & $\Omega_1(\infty)$ \\
\hline
1/N expansion & \protect{\cite{CSY}} & 0.2718 \\
classical Monte Carlo & \protect{\cite{CSY}} & $0.25\pm0.04$ \\
quantum Monte Carlo & this study & $0.26\pm0.01$\\
\end{tabular}
\end{table}

Chubukov, Sachdev and Ye have investigated the quantum critical regime
of the QNL$\sigma$M in close detail \cite{CSY}.  They make some
further predictions based on scaling arguments. On the ordered side
the spin stiffness $\rho_s$ vanishes as
\begin{equation}
\rho_s\propto(g_c-g)^{(d+z-2)\nu}= (g_c-g)^{\nu},
\end{equation}
where the second equivalence comes from the prediction that $z=1$.
Additionally it follows from general scaling arguments that the
uniform susceptibility at the critical point is universal:
\begin{equation}
\chi_u = \Omega_1(\infty)\left({g\mu_B\over \hbar c}\right)^2 T.
\end{equation}
Here $c$ is the spin wave velocity and
$\Omega_1(\infty)$ a universal constant. Estimates for $\Omega_1(\infty)$
are listed in Tab. II.

The spin wave velocity $c$ scales as
\begin{equation}
c\propto(g_c-g)^{\nu(z-1)}
\end{equation}
and is thus regular at the critical point if $z=1$.

\subsection{What about Berry phases?}

The equivalence of the 2D QAFM to the 2D QNL$\sigma$M however is still
an open question because of the existence of Berry phase terms in the
QAFM that are not present in the QNL$\sigma$M \cite{haldane}.  It has
been argued that these terms cancel in special cases, such as in the
bilayer model \cite{sandvik,Duin}.  Then it is plausible that the
quantum phase transition is in the same universality class as the
QNL$\sigma$M. This was confirmed by quantum Monte Carlo calculations
of Sandvik and coworkers
\cite{sandvik,sandvik2,sandvik3}. They have investigated the finite
size scaling of the ground state structure factor and susceptibilities
on lattices with up to $10 \times 10\times 2$ spins. Although these
lattices are quite small they still found good agreement of the
exponents $z$ and $\eta$ with the QNL$\sigma$M predictions
\cite{sandvik,sandvik2}.

In another study Sandvik {\it et al.} \cite{sandvik3} have
investigated finite temperature properties of the bilayer QAFM on
larger lattices and also found good agreement with the QNL$\sigma$M
predictions. In the absence of Berry phase terms the equivalence of
the QAFM and the QNL$\sigma$M is quite well established by these
simulations.

But in general these Berry phase terms exist. Chakravarty {\it et al.}
argue that they can change the critical behavior and lead to different
exponents \cite{CHN,Chakravarty}.  Chubukov {\it et al.} on the other
hand argue that the Berry phase terms are dangerously irrelevant
\cite{CSY} and do not influence the critical behavior.

Previous numerical simulations on dimerized square lattices
\cite{sandvik2,katoh} are indeed not consistent with the QNL$\sigma$M
predictions.  Sandvik and Veki\'c \cite{sandvik2} find a dynamical
exponent $z\ne1$, but their largest system was only $10 \times 10$
spins. The deviation could be a problem with scaling arising from
inequivalent spatial directions.

Katoh and Imada however found $z\approx1$, compatible with Lorentz
invariance. Additionally they calculated the correlation length $\xi$
and from it the exponent $\nu$. The validity of their result
$\nu\approx 1$ is however again questionable because because of the
restriction to very small lattices of $12\times 12$.  On the other
hand the discrepancy could be an effect of the Berry phase terms that
are present in the dimerized square lattice but probably not in the
bilayer.

The main purpose of the simulations reported is to she light onto this
question and to clarify the role of the berry phase terms. Our results
support the ideas of Chubukov {\it et. al.} \cite{CSY} that the Berry phase terms
are dangerously irrelevant.

\section{Algorithm and parallelization}

Using the new quantum cluster algorithms, the loop algorithms
\cite{evertz,beard} it is possible to simulate much larger lattices at
lower temperatures, just as the corresponding classical cluster
algorithms have allowed the simulation of critical classical spin
systems. With these algorithms it has for the first time become
possible to study quantum critical spin systems in detail.

A disadvantage of the cluster methods however is that they cannot be
vectorized as easily as the local update algorithms. Using powerful
vector machines is therefore not an option. Fortunately however most
of the modern supercomputers are parallel machines, and Monte Carlo
methods are nearly ideally suited for that architecture.

One of the authors has developed an object oriented Monte Carlo
library in C++ \cite{alea}. Using this library it is very simple to
parallelize a Monte Carlo program and to port it to new parallel
computers. The library automatically parallelizes any Monte Carlo
simulation at the two ``embarrassingly parallel'' levels. The first
level of trivial parallelism is the parameter parallelism. Simulation
with different parameters, such as system size, coupling or
temperature can be performed independently in parallel. At this level
there is practically no overhead due to the parallelization. We get
perfect speedup and the library takes care of load balancing.

A single simulation can similarly be parallelized by running it in
parallel with different initial states and random seeds on each of the
processors. The simulations run nearly independent. Communication is
required only at the start and the end of the simulation. This level
of parallelization incurs some overhead however.  The overhead is the
time used to thermalize a simulation. We loose efficiency if this
thermalization time becomes comparable to the time actually needed for
the simulation.

The third and deepest level of parallelization cannot be automatically
done by the library since it depends on the algorithm used for Monte
Carlo. The lattice used for one simulation can be spread over many
processors.  This parallelization has to be done by the programmer of
the algorithm, but it is supported by various functions of the
library.  It is worthwhile to invest time in this parallelization only
in two cases.  The first is when, as mentioned above, thermalization
is slow.  Often the main reason is however different one. Large
lattices simply might not fit into the memory of one processor.

In our simulations reported here we have used the 1024-node, 300 GFlop
Hitachi SR2201 massively parallel computer of the university of
Tokyo. At the time of its introduction this machine was the fastest
general purpose computer in the world. Each processor has 256 MByte of
local memory, enough to simulate quantum spin systems with 20000 spins
at temperatures as low as $T=0.01$. This was large enough for the
present study and we did not spend time on the third level of
parallelization but used only the first two levels provided by the
library.

The algorithm used was the continuous time loop algorithm
\cite{beard}.  The loop algorithms, first developed by Evertz et
al. \cite{evertz} are quantum version of the classical cluster
algorithms. The continuous time version is preferable over the earlier
discrete time versions since it eliminates the need to extrapolate in
the finite Trotter time step $\Delta\tau$. In our experience we found
that this leads to a four-fold speed increase. Additionally the
continuous time algorithm uses only 10\% of the memory compared to the
discrete time algorithm, allowing the simulation of larger lattices.

\section{The CaV$_4$O$_9$ lattice}

\begin{figure}
\begin{center}
\mbox{\psfig{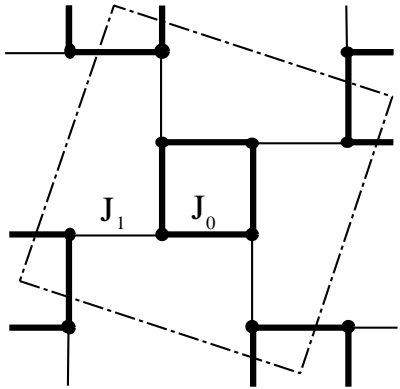}}
\end{center}
\caption{Lattice structure of the 1/5-th depleted square lattice
of CaV$_4$O$_9$. The dashed square indicates the eight spin unit cell
used in our calculations.}
\label{fig:lattice}
\end{figure}

As the universality class of a phase transition does not depend on the
microscopic details of the lattice structure we are free to choose the
best lattice for our purposes. We have chosen the ${\rm CaV}_4{\rm
O}_9$ lattice, a 1/5-th depleted square lattice depicted in
Fig. \ref{fig:lattice} for our calculations. There are three reasons
for this choice. Firstly the Berry phase terms are present on this
lattice \cite{Sachdev}. Next both space directions are equivalent, in
contrast to the dimerized square lattice \cite{sandvik2,katoh}. This
makes the scaling analysis easier. Finally at the quantum critical
point all the couplings are nearly equal in magnitude, which is also
optimal from a numerical point of view. We have performed our
simulations on square lattices with $N=8n^2$ spins, where $n$ is an
integer. Our largest lattices contained 20 000 spins. For the
following discussion it is useful to introduce the linear system size
$L$ in units of the bond lengths $a$ of the original square lattice:
$L\equiv \sqrt{5N/4} a$.

\begin{figure}
\begin{center}
\mbox{\psfig{figure=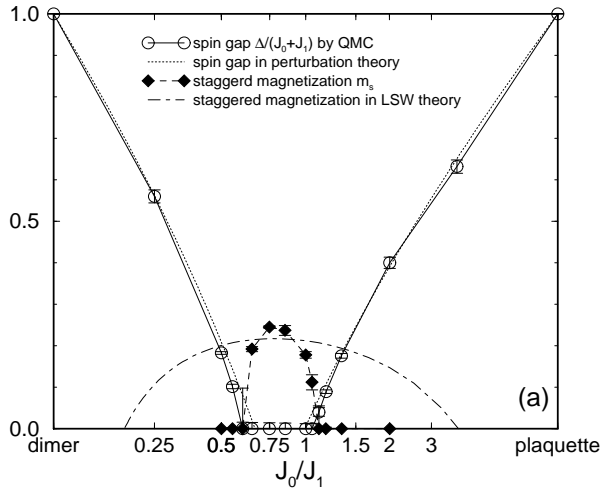,width=9cm}}
\end{center}
\caption{Phase diagram of the CaV$_4$O$_9$ spin lattice as a function of the
ratio $J_0/J_1$, reprinted from Ref. \protect{\cite{cvo}}. The leftmost
point corresponds to the dimer limit $J_0=0$ and the rightmost point
to the plaquette limit $J_1=0$. Circles indicate quantum Monte Carlo
results for the spin gap, normalized by $J_0+J_1$. Diamonds show the
staggered magnetization. As reference the perturbation theory
estimates for the gap\protect{\cite{ueda}} and the linear spin wave
theory (LSW) estimates for the staggered moment have been included.}
\label{fig:cvo}
\end{figure}

The phase diagram of this lattice has been discussed in detail in Ref.
\cite{cvo}  and is shown in Fig. \ref{fig:cvo}.
By removing every fifth spin we obtain a lattice consisting of
four-spin plaquettes linked by dimer bonds. We label the couplings in
a plaquette $J_0$ and the inter-plaquette couplings $J_1$.  By
controlling the ratio of these couplings $J_1/J_0$ we can tune from
N\'eel order at $J_1=J_0$ to a quantum disordered ``plaquette RVB''
ground state with a spin gap $\Delta=J_0$ at $J_1=0$.

\section{Results}
\subsection{The critical point}
The first step in the determination of the critical behavior is a high
precision estimate of the critical coupling ratio $J_1/J_0$. We have
calculated the second moment correlation length $\xi_L$ on systems of
various sizes $L$. This can be determined in the usual way from the
magnetic structure factor $S({\bf q})$ close to the N\'eel peak at
${\bf Q}$:
\begin{equation}
S({\bf Q}+\delta{\bf q}) = {S({\bf Q})\over1+(||\delta{\bf q}||\xi)^2}
+ {\rm O}(\delta{\bf q}^4).
\end{equation}

\begin{figure}
\begin{center}
\mbox{\psfig{figure=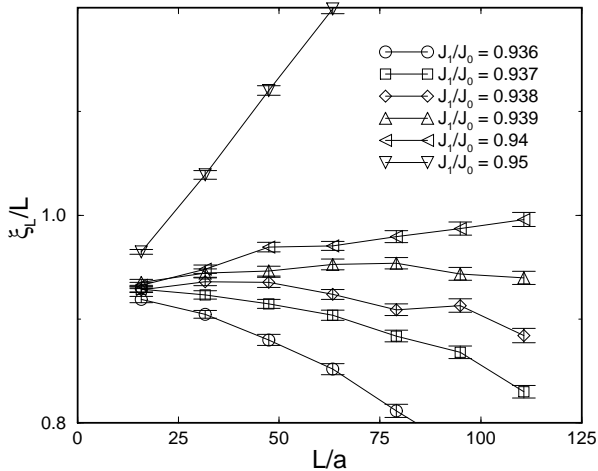,width=9cm}}
\end{center}
\caption{Plot of the ratio of correlation length divided by system size
$\xi_L/L$. At the critical point the correlation length calculated in a
finite system is proportional to the system size. This is the case for
$(J_1/J_0)_c=0.939\pm0.001$.}
\label{fig:cl}
\end{figure}

The temperature was chosen to be $T=J_0/L$, keeping the finite $2+1$
dimensional system in the cubic regime.  From standard finite size
scaling arguments it follows that this correlation length $\xi_L$
scales proportional to the system size $L$ at criticality. We have
calculated the ratio $\xi_L/L$ (shown in Fig. \ref{fig:cl}) for a
variety of couplings and system sizes up to $N = 9600$. Independence
of the system size was seen at the critical coupling ratio
$(J_1/J_0)_c=0.939\pm0.001$.

\begin{figure}
\begin{center}
\mbox{\psfig{figure=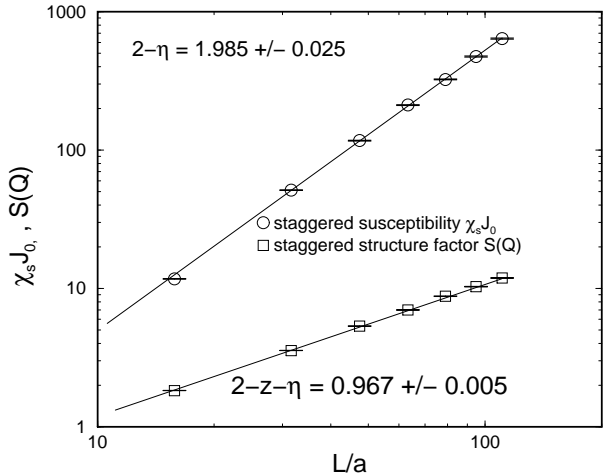,width=9cm}}
\end{center}
\caption{Finite size scaling of the staggered structure factor and
susceptibility at the critical point.}
\label{fig:fss}
\end{figure}

\subsection{The exponents}
Next we have calculated the finite size scaling of both the staggered
structure factor $S({\bf Q}) = L^2m_s$ and of the corresponding staggered
susceptibility. At criticality they scale like
\begin{eqnarray}
S({\bf Q})& \propto& L^{2-z-\eta}\\
\chi_{s}& \propto& L^{2-\eta}
\end{eqnarray}
The temperature was chosen to be $T=J_0/(6L)$, low enough to see ground
state properties within our accuracy. By fitting the results shown in
Fig. \ref{fig:fss} we obtain the estimates $z=1.018\pm0.02$ and
$\eta=0.015 \pm 0.020$. This is perfectly consistent with the Lorentz
invariance ($z=1$) expected from a mapping to the QNL$\sigma$M. We
will discuss $\eta$ below together with the other exponents. From
these fits it is also obvious that at least $N=800$ spins are
necessary to obtain good scaling.

The remaining exponents $\beta$ and $\nu$ are best calculated from the
magnetization $m_s$ and the spin stiffness $\rho_s$ on the ordered
side. Good estimates for $m_s$ and $\rho_s$ can be obtained by the
Hasenfratz-Niedermayer equations \cite{HN}. These authors have
calculated the {\it exact} finite-size and finite-temperature values
of the low-temperature uniform and staggered susceptibilities $\chi_u$
and $\chi_s$ for the ordered phase of a 2D QAFM on a lattice with the
symmetries of a square lattice.  Their equations for the staggered
susceptibility
\begin{eqnarray}
\chi_s(T,L)&= & \frac{{\cal M}_s^2 L^2}{3 T} \left\{1 +
2 \frac{\hbar c}{\rho_s L l}
\beta_1(l) + \left(\frac{\hbar c}{\rho_s L l}\right)^2 \left[\beta_1(l)^2 +
3 \beta_2(l)\right] + ...\right\}
\end{eqnarray}
and for the uniform susceptibility
\begin{eqnarray}
\chi_u(T,L) = \frac{2 \rho_s}{3 (\hbar c)^2} &&\left\{+ \frac{1}{3}
\frac{\hbar c}{\rho_s L l} \tilde{\beta}_1(l) + \right. \\
&&\qquad \left.\frac{1}{3} \left(\frac{\hbar c}{\rho_s L l}\right)^2
\left[\tilde{\beta}_2(l) - \frac{1}{3} \tilde{\beta}_1(l)^2 - 6 \psi(l)
\right] + ...\right\}. \nonumber
\end{eqnarray}

are correct for the low temperature regime $k_B T
\ll 2\pi\rho_s$ with cubic geometry $l^3 = {\hbar c\over T L} \approx 1$.
Up to second order in $T$ (or $1/L$ respectively) the susceptibilities
are universal, determined by only three parameters: the staggered
magnetization $m_s$, the spin stiffness $\rho_s$ and the spin wave
velocity $c$. The shape functions $\beta_1$, $\beta_2$,
$\tilde{\beta}_1$, $\tilde{\beta}_2$ and $\psi$ are known exactly for
square lattice geometries.  Two high precision quantum Monte Carlo
studies have confirmed the validity of these equations for the square
lattice QAFM
\cite{beard,wiese}.

\begin{figure}
\begin{center}
\mbox{\psfig{figure=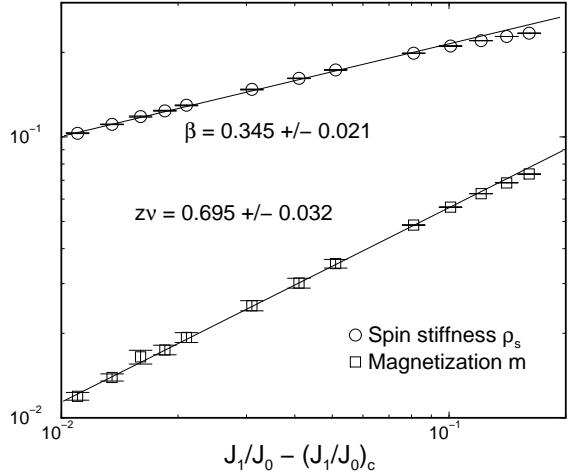,width=9cm}}
\end{center}
\caption{Staggered magnetization $m_s$ and spin stiffness $\rho_s$ calculated
by a fit of the low temperature susceptibilities on finite lattices to
the Hasenfratz-Niedermayer equations.}
\label{fig:hn}
\end{figure}

We have calculated the susceptibilities for a wide range of couplings
$0.95 < J_1/J_0 < 1.1$, lattice sizes $800 < N < 16200$ and
temperatures $0.006 < T/J_0 < 0.1$. The fits to the
Hasenfratz-Niedermayer equations are all excellent, with
$\chi^2/\mbox{\rm d.o.f.}\approx 1.5$.  This is another confirmation
of the universality of the Hasenfratz-Niedermayer equations.  From the
fits we obtain the staggered magnetization $m_s$, the spin stiffness
$\rho_s$ and the spin wave velocity $c$. The exponents $\beta$ and
$\nu$ can then be obtained in a straightforward way (see
Fig. \ref{fig:hn}) and are listed in Tab. I.

\subsection{Discussion}
Let us now discuss the results. First we observe that the exponents
satisfy the scaling relation Eq. (\ref{eq:scaling}), indicating the
validity of the scaling ansatz for this quantum phase transition. The
exponents $\beta$, $\nu$ and $\eta$ are in excellent agreement with
the exponents of the 3D classical O(3) or Heisenberg model. They are
however incompatible with the mean field exponents calculated by Katoh
and Imada on small lattices.

Assuming Lorentz invariance $z=1$ we can improve our estimates for the
other three exponents. The agreement of the improved estimates with
the 3D O(3) exponents becomes even better. We can also rule out the 3D
Ising universality class whose exponents are also listed in Table I
for a comparison.

This excellent agreement is a strong indication that the Berry phase
terms in the 2D QAFM are indeed dangerously irrelevant as suggested by
Chubukov, Sachdev and Ye. To further confirm their predictions we have
calculated the uniform susceptibility close to criticality down to
$T=0.02$, more than an order of magnitude lower than
Ref. \cite{sandvik}. We have extrapolated the finite size results on
lattices with up to $N=20 000$ spins to the thermodynamic
limit. Looking for the coupling at which a linear behavior occurs
gives an independent estimate of the critical point:
$(J_1/J_0)_c=0.939\pm0.002$, in excellent agreement with the above
estimate. The linear slope is $\Omega_1(\infty)(aJ_0/\hbar c)^2 =0.238
\pm 0.003 $.  By extrapolating the spin wave velocity determined in
the ordered phase by the Hasenfratz-Niedermayer fit to the critical
point we get $\hbar c/aJ_0=1.04\pm0.02$ and thus $\Omega_1(\infty) =
0.26\pm0.01$, again in excellent agreement with Chubukov et al. (see
Tab. II).

\section{Summary and Outlook}

To summarize, we have performed a large scale quantum Monte Carlo
simulation of a quantum phase transition in a planar
antiferromagnet. The new quantum cluster algorithms, in particular the
continuous time loop algorithm allow high precision simulation of
critical quantum systems.

The critical exponents that we have calculated (listed in Table I)
agree within our errors with the exponents of the classical 3D O(3) or
Heisenberg model.  The dynamical exponent $z=1.018\pm0.02$, consistent
with Lorentz invariance. This is compelling numerical evidence for the
conjecture that the quantum Heisenberg antiferromagnet is in the same
universality class as the 2D quantum nonlinear sigma model and the 3D
Heisenberg ferromagnet. The Berry phase terms that are present in the
Heisenberg antiferromagnet for non-integer spin do not seem to be
influence the critical behavior. This supports the conjecture by
Chubukov, Sachdev and Ye \cite{CSY} that they are dangerously
irrelevant.

While the accuracy achieved in the present simulation is remarkable
for a simulation of a quantum system it is not very good compared to
the best classical results that are an order of magnitude more
accurate.  If one wishes for higher accuracy the best approach could
be to generalize the histogram methods to quantum systems and to do a
finite size scaling study of cumulants, similar to the ones done in
the classical case \cite{expO3}. Even then we might not be able to
reach the same accuracy as in the classical simulations for two
reasons. Firstly we cannot measure the order parameter $m_s$ in a
quantum Monte Carlo simulation, but only its square $m_s^2$. Therefore
we can only calculate the cumulants of $m_s^2$, which will be less
favorable numerically.  Another difference between a classical and a
quantum system is that in the 3D classical system all three space
directions are equivalent.  In the (2+1)D quantum system on the other
hand the time direction is not equivalent to the space direction,
which makes a scaling analysis more complex.

\acknowledgements
We want to thank the Computer Center of the university of Tokyo for
giving us the ability to use their 1024-node massively parallel
Hitachi SR 2201 supercomputer. Being able to use this fast computer
has enabled us to perform the simulations reported here.  We also want
to thank K. Ueda, J.-K. Kim, D.P. Landau, S. Sachdev A.W. Sandvik and
U.-J. Wiese for interesting discussions. M.T. was supported by the
Japan Society for the Promotion of Science JSPS.


\end{document}